\documentclass[12pt]{iopart}
\usepackage{epsfig}

\newcommand\cpc[3]{{{\it Comput. Phys. Commun. }{\bf #1} (#2) #3}}

\newcommand\plb[3]{{{\it Phys. Lett. }{\bf B #1} (#2) #3}}

\newcommand\epjc[3]{{ \it Eur. Phys. J. } {\bf {C#1}} (#2) #3}

\begin{document}

\vspace{-4cm}
\begin{flushright}
  UR--1588\\
  ER/40685/938\\
  hep-ph/9911447\\
\end{flushright}

\title{On the Top Mass Reconstruction Using Leptons}

\author{G. Corcella\footnote{Talk
given at the UK Phenomenology Workshop on Collider Physics,
Durham, U. K., 19-24 September 1999.}}
\address{Department of Physics and Astronomy, University of Rochester\\
Rochester, NY 14627, U. S. A.}
\begin{abstract}
I discuss the feasibility of measuring the top quark mass by the using
of final states with leptons and $J/\psi$ at hadron
colliders. I also investigate the impact of matrix-element corrections
to the HERWIG simulation of top decays.
\end{abstract}
Top quark physics is presently one of the most interesting fields of 
investigation in both theoretical and experimental analyses.
In the next Tevatron RUN II and, ultimately, at the LHC, a large amount
of $t\bar t$ pairs will be produced, which will allow precision measurements
of the top quark properties and especially of its mass.

In $t\bar t$ events at hadron colliders, the final states are classified
according to the decay modes of the two $W$ bosons produced in the decays
$t\to bW$.
At the Tevatron RUN I, the top mass was determined by means of the lepton +
jets and the dilepton modes by both CDF and D\O\ collaborations, CDF also
considering the all-hadron mode.
The estimated average value for the top mass reads \cite{pdg}:
$m_t = 173.8\pm 3.5\pm 3.9\ {\mathrm{GeV}}$.

In this talk I discuss the method of reconstructing the top mass by
means of final states where the two produced $W$ bosons decay
leptonically, i.e. $W\to l\nu$, $l$ being either an electron or a muon, 
and the $B$ meson, coming from the hadronization of the
$b$ quark, decays into a state containing a $J/\psi$, eventually decaying
into a $\mu^+\mu^-$ pair. According to 
the LHC experimentalists \cite{avto}, this
is a favourite channel, with the estimated systematic error 
being no larger than 1 GeV. 
In three years of high luminosity $L=3\times 10^5\ {\mathrm{pb}}^{-1}$, 
about $3\times 10^3$ final states with well-identified 
leptons and $J/\psi$s are foreseen.
The idea is that one should relate the peak value of the distribution 
of the invariant masses $m_{J/\psi l}$ and $m_{\mu l}$
to the top mass by the using of Monte Carlo
simulations.

The suggested channel presents some advantages which make it quite suitable: 
the backgrounds are small and can be 
suppressed by setting cuts on the transverse momentum and rapidity 
of the final-state leptons; the effects of the initial- and final-state 
radiation are negligible
and can be easily taken into account; the spectra $W\to l\nu$ and $B\to 
J/\psi$ are well known.

\begin{figure}
\centerline{\resizebox{0.50\textwidth}{!}{\includegraphics{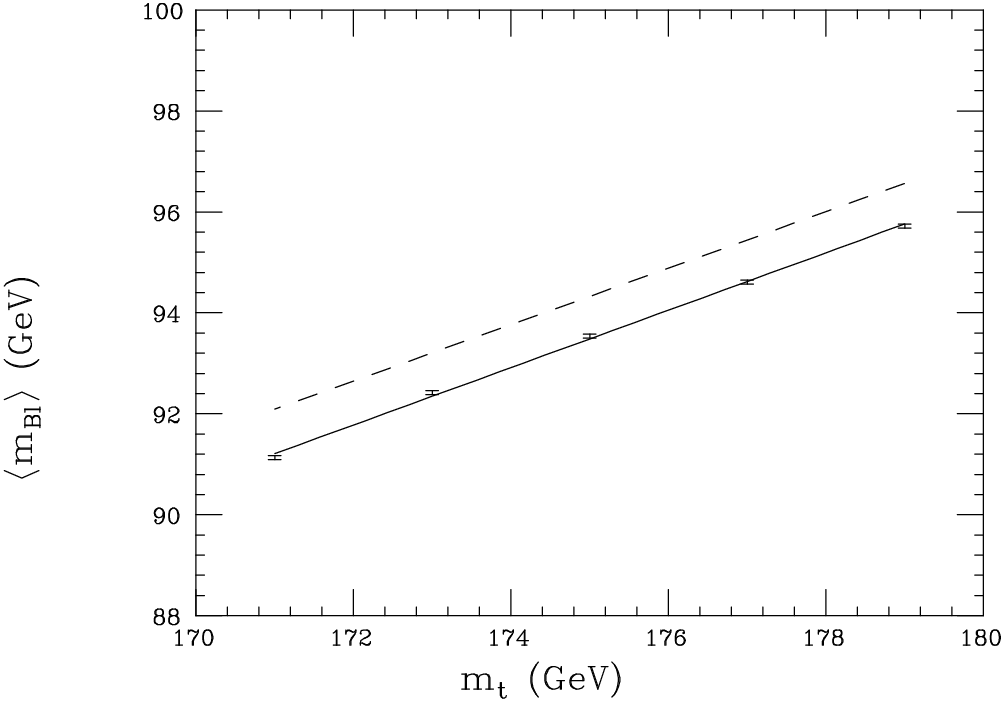}}}
  \caption{$\langle m_{Bl}\rangle $ as a function of
$m_t$ at the LHC after a fit into a straight line, according to
HERWIG 6.1 (solid line) and 6.0 (dotted).}
  \label{fitlhc}
\end{figure}

Since the proposed method crucially relies on the Monte Carlo model used to 
simulate top production and decay, 
I wish to investigate the effect of matrix-element 
corrections to top decays recently implemented \cite{corsey}
in the HERWIG algorithm \cite{herwig}.
As the $B\to J/\psi$ spectrum is well known, I shall consider the spectra
of $m_{Bl}$. In HERWIG 5.9, the latest public
version, a few bugs were found in the implementation of top decays.
These bugs are corrected in the intermediate version 6.0,
HERWIG 6.1 being the new version, including also matrix-element corrections
to top decays.
The Tevatron statistics will be too low to detect $J/\psi$s
via top quarks. It is however worthwhile performing the analysis even at 
the Tevatron to check the consistency of the method which claims 
to estimate the top mass independently of the 
production mechanism, which is mainly $q\bar q\to t\bar t$ at the Tevatron
and $gg\to t\bar t$ at the LHC.
\begin{figure}
\centerline{\resizebox{0.49\textwidth}{!}{\includegraphics{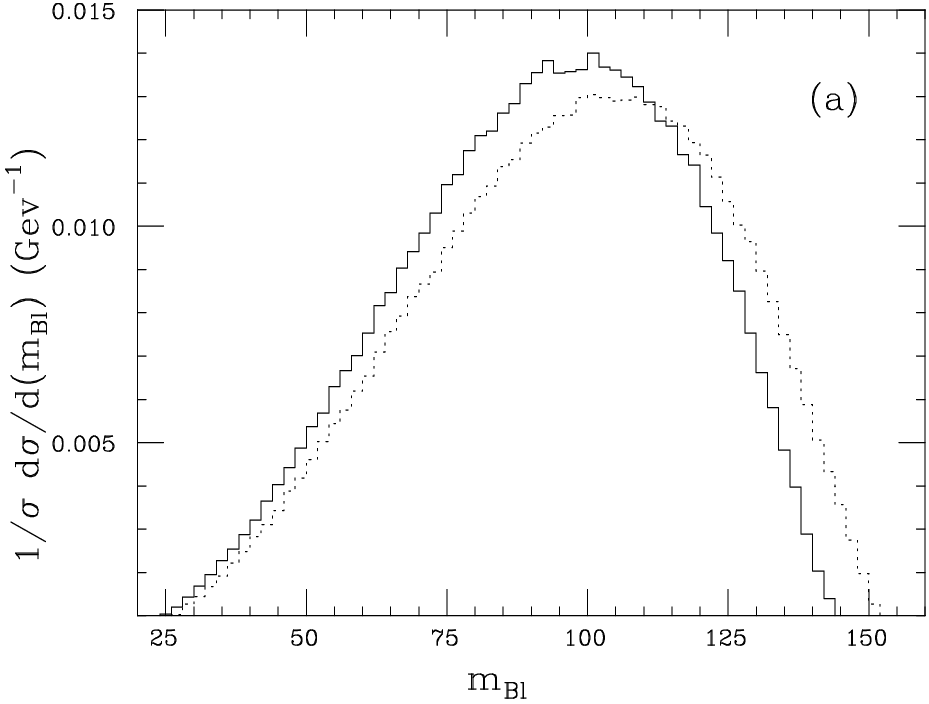}}%
\hfill%
\resizebox{0.49\textwidth}{!}{\includegraphics{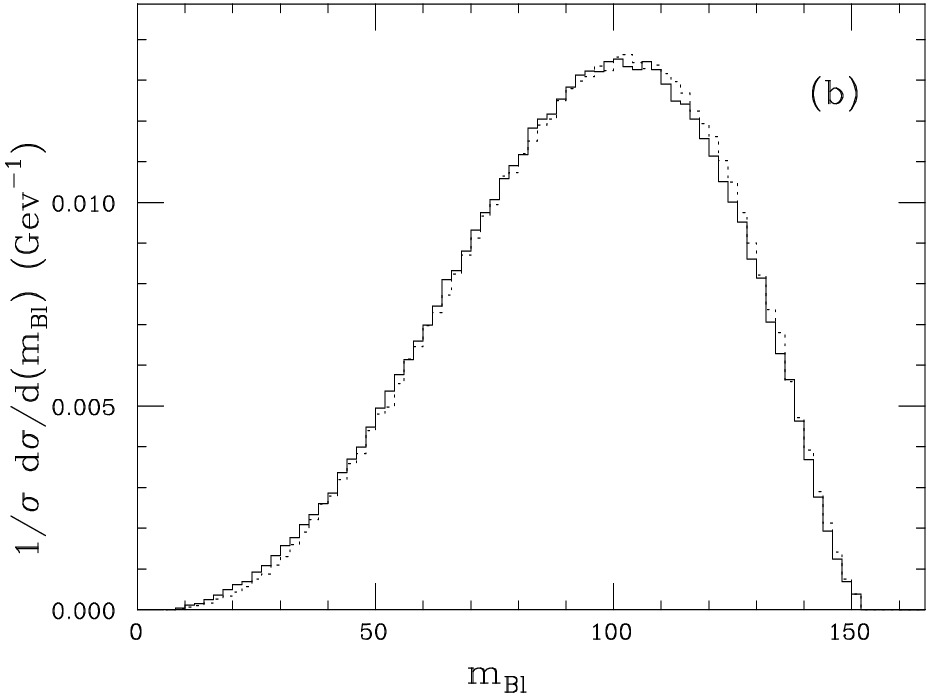}}}
  \caption{Invariant mass of the $B$-lepton system at the LHC for 
$m_t=171$~GeV (solid) and $m_t=179$~GeV (dotted), according to HERWIG 6.1 (a); 
for $m_t=175$~GeV, according to HERWIG 6.0 (dotted) and 6.1 (solid) (b).}
  \label{mass}
\end{figure}
In Fig.~\ref{mass} the invariant mass $m_{Bl}$ is plotted at the LHC; 
in tables \ref{masstev} and \ref{masslhc} 
one can find the average values $\langle m_{Bl}\rangle$ and the statistical 
errors on the obtained discrepancies for different values of the top mass.
We observe a systematic shift of about $800$~MeV towards lower values
of $\langle m_{Bl}\rangle$ after the inclusion of matrix-element corrections.
Also, the results at the Tevatron and at the LHC 
are the same within the range of 150 MeV.
If we parametrise the relation between $\langle m_{Bl}\rangle$ and 
$m_t$ as a straight
line, we find for the LHC, by means of a least square fit:
\begin{eqnarray}
6.1\ :\;
\langle m_{Bl}\rangle&=&0.568\ m_t-\ \,  6.004\ {\mathrm{GeV}}\ ,\ 
\epsilon=0.057\ {\mathrm{GeV}}\ ;\\
6.0\ :\;
\langle m_{Bl}\rangle&=&0.559\ m_t-\ \, 3.499\ {\mathrm{GeV}}\ ,\ 
\epsilon=0.052\ {\mathrm{GeV}}
\  ;\end{eqnarray}
where $\epsilon$ is the mean square deviation in the fit.
The straight line fits and the HERWIG 6.1 points in table \ref{masslhc}
are plotted in Fig.~\ref{fitlhc}:
from the values of the slopes, it follows that the 
discrepancies on $\langle m_{Bl}\rangle$ result in an impact of about 
$\Delta m_t\approx 1.5$~GeV on the top mass.

In summary, I considered the top mass reconstruction at the LHC 
using final states 
with leptons and $J/\psi$ and found that the effect of matrix-element
corrections to the HERWIG simulation of top decay is of about 1.5 GeV.
More details on this analysis will be found in \cite{cormansey}, where we shall
also study jet distributions and the $b$-quark fragmentation.
\begin{table}[t]
\begin{center}
\begin{tabular}{||l|r|r|r|r||}\hline
\ \ \  $m_t$\ \ \ 
&$\langle m_{Bl}\rangle ^{6.1}$&$\langle m_{Bl}\rangle ^{6.0}$
&$\langle m_{Bl}\rangle ^{6.0}-\langle m_{Bl}\rangle ^{6.1}$\\ \hline
171 GeV&91.18 GeV&92.06 GeV&$(0.873\pm 0.037)$ GeV\\\hline
173 GeV&92.31 GeV&93.22 GeV&$(0.912\pm 0.038)$ GeV\\\hline
175 GeV&93.41 GeV&94.38 GeV&$(0.972\pm 0.038)$ GeV\\\hline
177 GeV&94.65 GeV&95.45 GeV&$(0.801\pm 0.039)$ GeV\\\hline
179 GeV&95.64 GeV&96.63 GeV&$(0.984\pm 0.039)$ GeV\\\hline
\end{tabular}
\end{center}
\caption{Results at the Tevatron
for different values of $m_t$.\label{masstev}}
\end{table}
\begin {table}
\begin{center}
\begin{tabular}{||l|r|r|r|r||}\hline
\ \ \ $m_t$\ \ \ 
&$\langle m_{Bl}\rangle ^{6.1}$&
$\langle m_{Bl}\rangle ^{6.0}$&$\langle m_{Bl}\rangle ^{6.0}-\langle 
m_{Bl}\rangle ^{6.1}$\\ \hline
171 GeV&91.13 GeV&92.02 GeV&$(0.891\pm 0.038)$ GeV\\\hline
173 GeV&92.42 GeV&93.26 GeV&$(0.844\pm 0.038)$ GeV\\\hline
175 GeV&93.54 GeV&94.38 GeV&$(0.843\pm 0.039)$ GeV\\\hline
177 GeV&94.61 GeV&95.46 GeV&$(0.855\pm 0.039)$ GeV\\\hline
179 GeV&95.72 GeV&96.51 GeV&$(0.792\pm 0.040)$ GeV\\\hline
\end{tabular}
\end{center}
\caption{As in table 1, but for the LHC.\label{masslhc}}
\end{table}

\section*{Acknowledgements}
I acknowledge M.L. Mangano and M.H. Seymour for their help in obtaining 
the presented results.
I am also grateful to A. Kharchilava and B.R. Webber for useful suggestions.


\section*{References} 
\begin{thebibliography}{99}
\bibitem{pdg}
C. Caso et al., {\it Review of Particle Physics}, \epjc{3}{1998}{1}.
\bibitem{avto}
 A. Kharchilava, CMS note -- 1999/065.
\bibitem{corsey}
 G. Corcella and M.H. Seymour, \plb{442}{1998}{417}. 
\bibitem{herwig}
  G. Marchesini et al., \cpc{67}{1992}{465}.
\bibitem{cormansey} 
 G. Corcella, M.L. Mangano and M.H. Seymour, in preparation.
\end{thebibliography}
\end{document}